\documentclass[%
 reprint,
 amsmath,amssymb,
 aps,
]{revtex4-2}

\usepackage{graphicx}
\usepackage{dcolumn}
\usepackage{bm}
\usepackage{mathtools}
\usepackage{physics}
\usepackage{subcaption}
\usepackage{comment}
\usepackage{cancel}
\usepackage{amsmath} 
\usepackage{cleveref}
\usepackage[inline]{enumitem}

\begin{document}

\preprint{APS/123-QED}

\title{Normal modes and shockwaves in cold atoms}

\author{Francisco Raposo}
\thanks{francisco.valerio.raposo12@gmail.com}
\author{Hugo Terças}%
\affiliation{%
Golp/Instituto de Plasmas e Fusão Nuclear, Instituto Superior Técnico, Av Rovisco Pais 1 1049, Lisbon, Portugal
}

\date{\today}
\begin{abstract}
Numerical methods are developed to simulate the dynamics of atoms in a Magneto-Optic Trap (MOT), based on the fluid description of ultracold gases under laser cooling and magnetic trapping forces. With this model, equilibrium hydrostatic profiles and normal modes are calculated, and numerical results are validated against theoretical predictions. As a test case, shock wave formation due to rapid gas expansion and contraction
of the ultracold gas is simulated. The latter is caused by a sudden change of the value of its effective collective charge. Limitations of the current methods and future improvements are discussed. This work provides a foundation for studying numerically complex MOT behaviors and their use as analog simulators for astrophysical phenomena.
\\
\end{abstract}

\maketitle

\section{Introduction}
\label{sec:intro}
The achievement of laser cooling and atom trapping in a experimental setting is, in the context of scientific history, a relatively recent event.
It emerged in the late twentieth century, following the invention of the laser and major advances in our understanding of
quantum phenomena, in particular the interaction between light and matter. This progress led to the concept of
laser cooling and the creation of the Magneto-Optic Trap (MOT) ~\cite{phillips2002nobel}.
This achievement has had lasting impacts on various areas of research, most notably in cold-matter physics.
Key developments, such as the realization of the Bose-Einstein Condensate (BEC) in a laboratory setting
and the refinement of atomic clocks, are direct beneficiaries of this technology.
In recognition of their pioneering work in laser trapping, William Phillips, Claude Cohen-Tannoudji,
and Steven Chu were awarded the Nobel Prize in Physics in 1997.

Theoretical insights and experimental innovations in this area have since allowed for increasingly
precise control over these systems, extending their applications well beyond their original purpose.
One exciting avenue opened by this flexibility is the use of MOTs as physical
simulators for systems with analogous dynamics, such as stars. This idea is driven by
similarities in the fundamental behaviors involved. Thus, the MOT offers a controlled
environment for studying complex astrophysical phenomena, including gravitational collapse
and the formation of shock waves, within the limits of valid analogies. This helps overcome one
of the major obstacles in fields like astrophysics: the vast distances involved and the lack of experimental control
over celestial objects.

To explore such phenomena, one can develop numerical methods as a foundation for deeper analytical and experimental
research. These methods must account for the nonlinear nature of the system and address potential numerical instabilities.
By comparing results against established benchmarks, we can validate the effectiveness of our models and
build confidence in future investigations.

This paper is organized as follows:
Section I – Introduction;
Section II – The Magneto-Optic Trap (MOT);
Section III – Hydrodynamic Model for the Ultra-Cold Atomic Gas;
Section IV – Numerical Simulation;
Section V – Numerical Results.


\section{The Magneto Optic Trap (MOT)}
\label{sec:MOT}
Magnetic traps have been employed to confine neutral atoms since 1985~\cite{PhysRevLett.54.2596}, utilizing magnetic field gradients to produce potential minima. The \textbf{Magneto-Optical Trap (MOT)} enhances this mechanism by incorporating optical molasses~\cite{PhysRevLett.57.314, PhysRevLett.55.48}, which is based on laser cooling. The simplest MOT setup comprises six counter-propagating laser beams with opposite circular polarizations ($\sigma_{+}$, $\sigma_{-}$), aligned along three orthogonal axes, and a spatially varying magnetic field generated by anti-Helmholtz coils~\cite{atoms10030083}.

In this configuration, the lasers decelerate and cool the atoms via radiative pressure, while the magnetic field provides a position-dependent restoring force. The resulting total force is given by

\begin{equation}
\vb{F}_{\text{MOT}} \approx -\eta \vb{v} - \kappa \vb{x},
\end{equation}

where $\kappa$ is the effective spring constant of the trap and $\eta$ is the damping coefficient, making the system analogous to an overdamped harmonic oscillator.

This restoring force can also be expressed through an effective potential:

\begin{equation}
\phi_{\text{MOT}} = \frac{1}{2} m \omega_{0}^2 \vb{x}^2,
\end{equation}

where $\omega_{0} = \sqrt{\kappa / m}$ denotes the trapping frequency.

For atom numbers $N \gtrsim 10^5$ and at high densities, multiple photon scattering becomes significant, introducing two additional collective effects: a repulsive radiation pressure force $\vb{F_R}$ and an attractive attenuation force due to laser absorption, $\vb{F_A}$~\cite{SeskoCollectiveForces, DALIBARD1988203}. These forces can be described by Poisson-like equations, and their sum,
$\vb{F}_C=\vb{F}_A+\vb{F}_R$
results in a net force satisfying
\begin{equation}
\label{Poisson1}
\div \vb{F}_C = Q,n,
\end{equation}
where $Q = \sigma_L ( \sigma_R - \sigma_L) I/c$. Here, $\sigma_R$ and $\sigma_L$ denote the scattering and absorption cross sections, respectively, $I$ is the unattenuated light intensity, and $c$ is the speed of light. Since $\sigma_R > \sigma_L$, we have $Q > 0$, indicating a repulsive effective interaction. The parameter $Q$ behaves analogously to the square of an effective atomic charge, drawing a formal resemblance to Coulomb interactions.

The strength of these collective effects is characterized by the effective plasma frequency $\omega_P^2 = Q n(0)/m$. As with the trapping force, this interaction derives from a potential $\phi_C$, defined via $-\boldsymbol{\nabla} \phi_C = \vb{F}_C$, leading to
\begin{equation}
\nabla^2 \phi_C = - Q n.
\end{equation}
For $N < 10^5$, thermal motion dominates, and the atomic cloud follows a Maxwell-Boltzmann distribution with characteristic width $\left({k_B T}/{m \omega_0^2}\right)^{1/2}$. At higher densities, multiple scattering becomes dominant, leading to a nearly uniform “waterbag” density profile ~\cite{Romain_2010}.

The dynamics of the gas are described by the hydrodynamic equations obtained from the zeroth and first moments of the Vlasov equation~\cite{PhysRevA.78.013408, Tercas-Physics-of-Ultra-Cold-Matter}:

\begin{equation} ~ \label{Eq: FluidModel}
\begin{dcases}
\frac{\partial n}{\partial t} + \div{\left(n \vb{v}\right)} = 0 \\
\frac{\partial \vb{v}}{\partial t} + \vb{v} \cdot \grad{\vb{v}} + \frac{\grad{P}}{m n} = \frac{\vb{F}}{m}
\end{dcases},
\end{equation}

where $n(t,\vb{r})$ is the number density, $P(t, \vb{r})$ is the pressure, and $\vb{F}$ is the total external force.

In many contexts, it is convenient to recast these equations in terms of the mass density $\rho(t,\vb{r}) = m n$ and the momentum density $\vb{p} = m \vb{v}$:

\begin{equation}
\begin{dcases}
\frac{\partial \rho}{\partial t} + \div{\vb{p}} = 0 \
\frac{\partial \vb{p}}{\partial t} + \frac{\vb{p}}{\rho} \\\div{\vb{p}} + \vb{p} \cdot \grad{\left(\frac{\vb{p}}{\rho}\right)} + \grad{P} = \frac{\rho}{m} \vb{F}
\end{dcases}
\label{Eq: FluidModelP}
\end{equation}

Assuming spherical symmetry, all quantities are taken to depend only on the radial coordinate ($n = n(r)$, $\vb{v} = v(r) \hat{e}_R$, etc.), where $r = \abs{\vb{x}}$ denotes the distance from the origin. Under this assumption, the hydrodynamic equations reduce to:

\footnote{Throughout this text, partial derivatives are abbreviated as $\partial_{x}$ for clarity and brevity.}

\begin{equation}
\begin{dcases}
\partial_{t} \rho + r^{-2} \partial_{r}\left(r^2 p \right) = 0 \\
\partial_{r} p +
r^{-2} \partial_{r}\left(r^{2} \frac{p^{2}}{\rho} \right)
+ \partial_{r} P = \frac{\rho}{m} F
\end{dcases}
\label{Eq: FluidModelPSpherical}
\end{equation}

where

\begin{equation}
F = - \kappa r - \eta v + F_C \quad\mathrm{and}\quad r^{-2}{\partial_r}\left(r^2 F_C\right) = Q,n.
\end{equation}

Analogous to the Lane-Emden model used in stellar structure theory~\cite{stellarStucEvol}, a polytropic equation of state is often adopted for ultra-cold trapped gases~\cite{PhysRevA.93.023404}:

\begin{equation}
P = C_{\gamma} n^\gamma,
\label{PolyPressure}
\end{equation}

where $C_{\gamma}= P(0)/n(0)^{\gamma}$ encodes the thermodynamic properties of the gas. This approach is motivated by the similarity in confinement mechanisms: an attractive central force balanced by a spatial pressure gradient. However, unlike self-gravitating systems, cold atomic clouds are confined by externally imposed forces independent of the density distribution. Moreover, multiple scattering introduces a repulsive force analogous to Coulomb interactions in plasmas. The polytropic index $\gamma$ captures the influence of these effects, with $\gamma = 1$ corresponding to the isothermal ideal gas limit. Further theoretical and experimental discussion is provided in~\cite{PhysRevA.93.023404}.
\subsection{Equilibrium profiles}
In hydrostatic equilibrium ($\vb{v} = 0$), the equilibrium density satisfies
\begin{equation}
    \frac{C_{\gamma}}{m} \frac{\gamma}{\gamma - 1} r^{-2} \frac{\partial}{\partial r}  
    \left(r^2 \frac{\partial n^{\gamma-1}}{\partial r} \right) = - 3 \omega_0^2 + \frac{Q}{m} n.
\end{equation}
In the isothermal limit, $\gamma \rightarrow 1$ and $Q = 0$, we recover
$C_1 = k_B T$, leading to
\begin{equation}
    n(r) = n(0) e^{- r^2 / 2 \sigma_T^2} \quad\mathrm{with}\quad 
    \sigma_T = \left(\frac{k_B T}{m \omega_0^2}\right)^{1/2},
    \label{Eq: isothermalregimequilibrium}
\end{equation}
implying that the cloud radius is primarily determined by the gas temperature.

If $Q = 0$ but the gas is non-ideal ($\gamma \neq 1$), then
\begin{equation}
    n(r) =  n(0) \left(1 - \frac{\gamma-1}{\gamma} 
    \frac{m \omega_0^2 n(0)}{P(0)} \frac{r^2}{2}\right)^{1/(\gamma-1)}.
\end{equation}
Note that as $\gamma \rightarrow 1$, the isothermal case is recovered.
In the multiple scattering regime, where $\omega_p \approx \sqrt{3} \omega_0$, the density profile becomes a water-bag model:
\begin{equation}
    n(r) = \frac{Q}{3 m \omega^2} n(0) \Theta(x - R),  
\end{equation}
with $\Theta$ being the Heaviside step function.
The general case is described by
\begin{equation}
    \frac{\gamma}{\gamma-1} \frac{1}{\xi^2}\frac{\partial}{\partial \xi} 
    \left(\xi^2\,\frac{\partial \theta }{\partial \xi}  \right) =
    -1 + \Omega_P^2 \theta^{1/(\gamma-1)},
    \label{eq:ThetaEquation}
\end{equation}
where $\theta = \left({n(r)}/{n(0)}\right)^{\gamma-1}$ and $\xi = r/R_\gamma$. Here, 
$R_\gamma = \left({P(0)}/{3 m \omega_0^2 n(0)}\right)^{1/2}$ is the polytropic radius, and
$\Omega_P^2 = {\omega_P^2}/{\omega_0^2}$ measures the ratio between the effective plasma and trap frequencies. 
Equation~\eqref{eq:ThetaEquation} generalizes the Lane-Emden equation, which governs hydrostatic equilibrium in stellar polytropes~\cite{PhysRevA.88.023412}. Notably, $\Omega_P^2$ is bounded between 0 and 1.

\subsection{Dimensionless equations}

The system’s characteristic timescale is $\tau \approx 1/\omega_0$, akin to a spring-mass oscillator's period. We define dimensionless variables as
$t^* = t\omega_0$, $\vb{x}^* = \vb{x}\omega_0/v_0$, 
$\vb{v}^* = \vb{v}/v_0$, and $\rho^* = \rho/\rho_0$.
Consequently, pressure scales as $P^* = P/P_0 = \rho^{*\,\gamma}$.

Assuming $n_0 = n(0)$ and
\begin{equation}
    v_{0} = v_s(0) \gamma^{-1/2}
    = \left( \frac{P(0)}{\rho(0)} \right)^{1/2},
\end{equation}
where $v_s = \partial P / \partial \rho$ is the sound speed, the equations in \Cref{Eq: FluidModelPSpherical} become
\begin{equation}
    \begin{dcases}
        \frac{\partial \rho^*}{\partial t^*} + \frac{1}{r^{*\,2}} \frac{\partial}{\partial r^*}\left(r^{*\,2} p^*  \right) = 0, \\
        \begin{split}
        \frac{\partial p^*}{\partial t^*} & + 
        \frac{1}{r^{*\,2}} \frac{\partial}{\partial r^*}\left(r^{*\,2} \frac{p^{*\,2}}{\rho^*} \right)
        + \frac{\partial P^*}{\partial r^*} \\ & = - \rho^* r^* - \eta^* p^* + 3 \Omega_P^2 \, \rho^*  F_C^*
        \end{split}
    \end{dcases},
    \label{Eq: FluidModelSphericalAdimChoiceA}
\end{equation}
where $P_0 = C_\gamma \rho(0)^\gamma / m^\gamma$ and $\eta^* = {\eta}/{m \omega_0}$.
The characteristic length is $x_0 = 3 R_\gamma$.

\subsection{Normal modes and collective oscillations}

Considering small perturbations from equilibrium:
$n \rightarrow n_{\text{eq}} + \delta \rho$,
$\vb{v} \rightarrow \delta \vb{v}$,
and $\vb{F}_C \rightarrow \vb{F}_{\text{eq}} + \delta \vb{F}_C$,
where $n_{\text{eq}}$, $\vb{v}_{\text{eq}} = 0$, and $\vb{F}_{\text{eq}}$ denote the equilibrium fields.

Inserting these into the fluid equations for $\gamma \neq 1$ yields:
\begin{equation}
    \begin{split}
        &- {\partial_t^2} \delta n - \eta\,{\partial_t} \delta n  =  
    - \frac{C_\gamma \gamma}{m} \div  \left(n_{\text{eq}}^{\gamma-1} \grad{\delta n} \right)
    \\ &  + \frac{C_\gamma}{m} \frac{\gamma-2}{\gamma-1} \div    \left( \delta n \grad{n_{\text{eq}}^{\gamma-1}} 
    \right) + \div{\left(n_{\text{eq}} \delta F_C\right)},
    \end{split}
    \label{Eq: gammaLocalPertubations}
\end{equation}
which differs from \cite{PhysRevA.88.023412} due to the $(\gamma -2)$ term.

In the isothermal regime ($\gamma = 1$),
\begin{equation}
    \begin{split}
        - \frac{\partial^2}{\partial t^2} \delta n - \eta \frac{\partial}{\partial t} \delta n  & = 
    - \frac{k_B T}{m} \div  \left(n_{\text{eq}} \grad{\left(\frac{\delta n}{n_{\text{eq}}}\right)}  
    \right) \\ & + \div{\left(n_{\text{eq}} \delta F_C\right)}.
    \end{split}
    \label{Eq: gamma1LocalPertubations}
\end{equation}

\subsubsection{Radial isothermal oscillations}

When $\gamma = 1$ and $\omega_P^2 = 0$, 
Eq.~\eqref{Eq: gamma1LocalPertubations}, combined with Eq.~\eqref{Eq: isothermalregimequilibrium}, leads to
\begin{equation}
    \begin{split}
        & \left(\omega^2 + i \omega \frac{\eta}{m} - \omega_0^2 \frac{l (l+1)}{\zeta^{2}} \right) R(\zeta) = \\
        & \omega_0^2 \zeta \frac{d R (\zeta)}{d\zeta} 
        - \omega_0^2 \left(\frac{d^2 R (\zeta)}{d\zeta^2} + 
        \frac{2}{\zeta} \frac{d R(\zeta)}{d\zeta} \right),
    \end{split}
    \label{eq:RadialEquationGamma1Eigenvalue}
\end{equation}
where $\zeta = 3 R_\gamma r$ and ${\delta n}/{n_{\text{eq}}} = R(r) f(\theta, \phi)$.

The angular part satisfies
\begin{equation}
    \frac{1}{\sin{\theta}} \frac{\partial}{\partial \theta} 
    \left(\sin{\theta} \frac{\partial f}{\partial \theta} \right)
    + \frac{1}{\sin^2{\theta}} \frac{\partial^2 f}{\partial \phi^2} = - l (l+1) f,
\end{equation}
mirroring the role of the quantum angular momentum number $l$.

To ensure convergence of Eq.~\eqref{eq:RadialEquationGamma1Eigenvalue}, the frequency spectrum must satisfy
\begin{equation}
    \omega^2_{n,l} = \omega_0^2 (l + 2 n),
    \label{eq:normalmodesisothermal}
\end{equation}
with damping included via $\omega = \omega_r + i \omega_i$, where
$\omega_r^2 = \omega^2_{n,l} - \eta^2 / (4 m^2 \omega_0^2)$ and
$\omega_i = -\eta / (2 m \omega_0)$.

\subsubsection{Non-isothermal small clouds}

For $\gamma \neq 1$ and $\omega_P^2 = 0$, small clouds obey
\begin{equation}
    \begin{split}
         & \omega^{2} R(\zeta)  +  
    \frac{(\gamma - 1) \omega_0^2}{2 \zeta^2} \partial_\zeta 
    \left[\zeta^2 (1-\zeta^2) \partial_\zeta R(\zeta)\right] \\ 
    & - (\gamma-2) \omega_0^2 \frac{1}{\zeta^2} \partial_\zeta 
    \left( \zeta^3 R(\zeta) \right) \\ 
    & - \frac{l (l+1) (\gamma - 1) \omega_0^2}{2 \zeta^2} (1-\zeta^2) R(\zeta) = 0,
    \end{split}
\end{equation}
with $\zeta = \sqrt{3} (\gamma-1)/(2 \gamma R_\gamma)\, r$.

Convergent solutions exist only if
\begin{equation}
    \begin{split}
        \omega_n^2 = & -\frac{1}{2} \omega_0^2 (\gamma - 1 ) l (l+1)  +
    \frac{1}{2} \omega_0^2 (\gamma - 1) (2 n + l) (2 n + 3) \\ 
    & - \omega_0^2 (\gamma - 2) (2 n + l + 3),
    \end{split}
    \label{eq:normalmodesmycasenotisothermal}
\end{equation}
which contrasts with \cite{PhysRevA.88.023412}, where
\begin{equation}
    \omega^2 = \omega_0^2 \left[l + 2 n \left((\gamma-1)(2 n + l + \tfrac{1}{2}) + 1\right)\right].
    \label{NormalModesGamma1Radial}
\end{equation}

For $\gamma = 1$, Eq.~\eqref{eq:normalmodesmycasenotisothermal} reduces to Eq.~\eqref{eq:normalmodesisothermal}. For $\gamma = 2$ (no angular motion), $\omega^2 = n(2n+3)\omega_0^2$ describes low-energy breathing modes in a collisionless Bose-Einstein condensate without kinetic pressure~\cite{RevModPhys.71.463}.
Unlike polytropic stars, which require $\gamma > 4/3$ for stability~\cite{Pinsonneault_Ryden_2023}, this model lacks such a condition. Additionally, the homologous expansion or collapse seen in $\gamma = 4/3$ stellar polytropes is not observed in ultracold gases.

\subsection{Shockwaves}

Nonlinearity in the system enables the formation of shock waves through wave breaking. Considering a frame where a discontinuity element is stationary and tangential velocities are continuous, the Rankine–Hugoniot conditions apply~\cite{landauVol6}:
\begin{align}
    \rho_1 v_1 &= \rho_2 v_2 \equiv j,
    \label{RankineHugoniot1} \\
    \rho_1 v_1^2 + p_1 &= \rho_2 v_2^2 + p_2,
    \label{RankineHugoniot2} \\
    \frac{1}{2} \rho_1 v_1^2 + w_1 &= \frac{1}{2} \rho_2 v_2^2 + w_2,
    \label{RankineHugoniot3}
\end{align}
where $\rho_i$, $v_i$, and $p_i$ are the density, velocity (relative to the shock), and pressure on either side ($i = 1$: front, $2$: back). The enthalpy density is $w_i = \rho_i + P_i \epsilon_i$, with $\epsilon_i$ the internal energy.

These express conservation of mass, momentum, and energy across the shock. Entropy must increase ($s_2 > s_1$), indicating irreversibility. Shock formation requires $v_1 > c_1$ and $V_1 > V_2$, indicating compression.

\section{Numerical Methods}
\label{sec:numerical}

Although the model is relatively simple, its non-linearity hinders the derivation of analytical solutions. Numerical methods therefore provide a flexible and robust framework for exploring its dynamics.

\subsection{Conservative equations}
As seen in \Cref{Eq: FluidModel}, the system resembles a conservative equation with additional trapping and collective force terms. In one spatial dimension,\footnote{This generalizes to multiple dimensions ($\vb{x} \in \mathbb{R}^n$).} the conservative form is written as:
\begin{equation}
{\partial_t} \vb{u}(x, t) + {\partial_x}{\vb{f}(\vb{u}(x, t))} = 0,
\label{conservativeequation}
\end{equation}
where $\vb{u}$ is an $n$-dimensional state vector of conserved quantities (e.g., mass, momentum, energy), and $\vb{f}$ is the corresponding flux. Conservation implies that $\int_L \vb{u} , dx$ over a domain $L$ remains constant in time.
If the Jacobian $\partial \vb{f} / \partial \vb{u}$ has real, distinct eigenvalues and a complete set of eigenvectors, the system is said to be \textbf{hyperbolic}. In our case, the flux $\vb{f}$ is nonlinear in $\vb{u}$, which complicates numerical solutions—particularly in the presence of discontinuities such as shock waves. Standard linear finite difference methods may fail under these conditions due to instability or convergence to non-physical solutions~\cite{leveque1992numerical}.

\subsection{Discretisation}To solve these equations, we define the mesh width
$\Delta x$ and the time step $\Delta t$,
such that the discrete points of the mesh
are given by $x_j = j \Delta x$ and $t_n = n \Delta t$,
where $j = \dots, -1, 0, 1, \dots$ and $n = 0, 1, 2, \dots$.
Moreover, the pointwise values of the true solution at $(x, t) = (x_j, t_n)$ are defined as
\begin{equation}
u_{j}^n = u(x_j, t_n),
\end{equation}
whereas the approximations to this solution, obtained numerically, are denoted by $U_j^n$.
When discussing the Lax-Friedrichs method, as in Section~\ref{sec:laxmethodsubsec},
$U_j^n$ is considered to be an approximation of the cell average:
\begin{equation}
U_j^n \coloneq \bar{u}{j}^n = \frac{1}{\Delta x}\int{x - \Delta x/2}^{x + \Delta x/2} u_j^n ,\text{d}x.
\end{equation}

\subsection{The Lax-Friedrich method}
\label{sec:laxmethodsubsec}To ensure the method is conservative, it must be written in conservation form%
 \cite{leveque1992numerical}. A classic example is the Lax-Friedrichs method, whose update rule is given by
\begin{equation}
\begin{split}
U_j^{n+1} &= \frac{1}{2} \left(U_{j+1}^n + U_{j-1}^n\right)
+ \frac{\Delta t}{2\Delta x} \left(f(U_{j+1}^n) - f(U_{j-1}^n)\right).
\end{split}
\end{equation}
This scheme advances the solution in time using spatial averaging, introducing artificial dissipation through numerical diffusion. It is first-order accurate in both time and space. In conservative form, the Lax-Friedrichs numerical flux is expressed as
\begin{equation}
\begin{split}
F(U_j, U_{j+1}) &= \frac{\Delta x}{2\Delta t} (U_j + U_{j+1})
+ \frac{1}{2} \left(f(U_j) + f(U_{j+1})\right).
\end{split}
\end{equation}

This flux is Lipschitz continuous and therefore consistent, ensuring local accuracy near discontinuities despite the presence of numerical diffusion or smearing, which is characteristic of the Lax-Friedrichs scheme. According to the Lax-Wendroff Theorem, if the numerical solution converges as $\Delta x, \Delta t \rightarrow 0$, it approaches a weak solution of the governing equations. To ensure physical validity, an entropy condition must also be satisfied. For gaseous systems such as the one considered here, this condition corresponds to the physical law of increasing entropy and is enforced through the energy equation within the Rankine-Hugoniot conditions. This yields a unique solution for variables such as $\rho$, $P$, and $v$ across discontinuities.

\subsection{Stability}
For the method at hand,
in order to be stable, it is necessary, but not sufficient, that~\cite{leveque1992numerical, Chen2001}
\begin{equation}
\frac{\Delta t}{\Delta x} \abs{\frac{\partial f}{\partial \vb{u}}} \leq 1.
\end{equation}
Here, the $\abs{\cdot}$ operator denotes the determinant. For the Euler fluid equations, this condition is equivalent to
$C = {\Delta t}/{\Delta x} \leq {1}/{\abs{\vb{v}}_{\text{max}}}$
where $\abs{\vb{v}}_{\text{max}}$ is the maximum absolute velocity, and $C$ is known as the Courant number.
This is referred to as the Courant–Friedrichs–Lewy (CFL) condition, which restricts the numerical propagation speed: all perturbations must travel slower than a certain threshold to avoid non-physical artifacts. While it imposes an upper bound on $C$, its precise value affects numerical dispersion—suppressing fast waves may distort high-frequency components and the overall solution profile.

\subsection{Source Terms and Spherical Symmetry}
\label{subsec:SourceTermsandSphericalSymmetry}

So far, we have considered only purely conservative equations in one dimension. However, as previously mentioned, this does not apply to our ultra-cold gas model. A more general form of the conservation equations includes a source term $\vb{S}(\vb{u}, x, t)$:
\begin{equation}
\partial_t \vb{u}(\vb{x}, t) + \div \vb{F}(\vb{u}(\vb{x}, t)) = \vb{S}(\vb{u}, \vb{x}, t),
\label{conservativeequationsourceterms}
\end{equation}
where $\vb{F}$ is the flux matrix.
If we also consider spherical coordinates, the atomic gas model can be expressed as
\begin{equation}
r^2 \partial_t \vb{u}(r, t)
+ r^2 \partial_r \vb{F}(\vb{u}(r, t))
= r^2 \vb{S}(\vb{u}, r, t),
\end{equation}
where
\begin{equation}
    \begin{split}
        & \vb{u} = 
    \begin{bmatrix}
        \rho \\
        p
    \end{bmatrix}, \, 
    \vb{F} = 
    \begin{bmatrix}
        p \\
        {p^2}/\rho  + P  \\ 
    \end{bmatrix},\,
    \\ & \vb{S} = \begin{bmatrix}
        0 \\
        \rho/m (- \kappa r - \eta {p}/{\rho} + F_C) + {2}/{r} P
    \end{bmatrix}.
    \end{split}
    \label{eq:conservativeequationterms}
\end{equation}
In this form, the Lax-Friedrichs method becomes:
\begin{equation}
    \label{eq:LAXFRIEDRICHSchemeSource}
    \begin{split}
        U_j^{n+1} & = {r_j^{-2}} / 2 
    \left(r_{j+1}^2 U_{j+1}^{n} + r_{j-1}^2 U_{j-1}^{n}\right) 
    \\ & + C r_j^{-2} / 2 \, \left( r_{j+1}^2 f(U_{j+1}^{n}) 
    - r_{j-1}^2 f(U_{j-1}^{n})\right) \\ +  & {\Delta t}{r_j^{-2}} / 2 \, \left( r_{j+1}^2 S^n_{j+1} 
    + r_{j-1}^2 S^n_{j-1} \right).
    \end{split}
\end{equation}

\subsection{Simulating the Ultracold Gas}
\label{sec:IMPLEMENTATIONLAXFRIEDRICH}

In order to simulate this, 
we consider that our space (grid) 
is divided
in $N$ grid points separated by $\Delta r = L /(N-1)$, 
($[\Delta r, L + \Delta r]$). The origin is not covered, because
spherical coordinates are being used.
Considering this discrete space, we apply the Lax-Friedrich 
scheme on the dimensionless equations. 
The time step is set by $\Delta t = C \Delta r$. The value of $C\leq 1 / v_{\text{max}}$ must be chosen as 
to maintain the stability of the calculations and depends on the kind of phenomena we want to study.

\subsection{Boundary conditions}

This method requires that we set boundary conditions on both frontiers of the domain. 
At the centre there is a singularity and as such cannot be directly calculated. 
One can extrapolate the value by considering ${\partial \rho}/{\partial r} (r = 0) = 0$.
Considering the $1^{\text{st}}$ order finite difference approximation of the derivative, this yields
$\rho_0^n = \rho_1^n$
The gas is always static on the central point, such that $p_0^n = 0$. On the other hand,
${\partial_r \rho} (r \rightarrow \infty) = 0$ and 
${\partial_r p} (r \rightarrow \infty) = 0$.
Equivalently, after discretising,
$\rho_N^n = \rho_{N-1}^n$ and $p_N^n = p_{N-1}^n$.

\subsection{Collective Force - The Fourier Transform Method}
\label{sec:CollectiveForce}

The equation for the collective force, in its dimensionless form, resembles
\footnote{The asterisks have been dropped, 
    for the sake of simplity}
\begin{equation}
    \nabla^{2} \phi_C = - n.
    \label{PoissonAdimChap3}
\end{equation}
By defining the Fourier Transforms (FT) of $n$ and $\phi_C$,
$\hat{n}(\vb{k})$ and $\hat{\phi}_C(\vb{k})$ \footnote{The convention used in this paper is
$\hat{f}(\vb{k}) = \int e^{-2 \pi i \vb{k} \cdot \vb{x}} f (\vb{x})\,\text{d}\vb{x}^{3}$ for the FT and
$f(\vb{k}) = \int e^{2 \pi i \vb{k} \cdot \vb{x}} \hat{f} (\vb{k})\,\text{d}\vb{k}^{3}$ for the IFT}, 
respectively, in k-space, \Cref{PoissonAdimChap3} becomes
\begin{equation}
   \hat{\phi}_C = {\hat{n}}/{ 4 \pi^2 \vb{k}^2 }, 
\end{equation}
which after considering spherical symmetry simplifies to
\begin{equation}
    \phi_C^{}(r) = \frac{1}{2 \pi^2 r} \int_0^{\infty} \sin{\left(2 \pi k r\right)} \frac{\hat{n}(k)}{k}  \,\text{d}k 
\label{CollectiveForceFourier}
\end{equation}
where
\begin{equation}
\hat{n}(k) = \frac{2}{k} \int_0^{\infty}\sin{\left(2 \pi k r^{'}\right)}\,r^{'} \, n(r^{'}) \,\text{d}r^{'}.
\label{SineFourierTransform}
\end{equation}
In practise, in the last few calculations we have simplified the three dimensional Fourier Transform 
into a much simpler Fourier Sine Transform. The final step to calculating the 
Collective Force is to perform the gradient, 
which, considering the spherical symmetry, is $F_C = - \partial_r \phi_C$.
\section{Summary of the method}
Overall, the method goes as follows: Starting from some initial conditions $U^0_j$, it calculates the values of
$U^n_j$ via the algorithm
\begin{enumerate*}[label=(\roman*),before=\unskip{ i.e., }, itemjoin={{, }}, itemjoin*={{, and }}]
    \item  At instant $t = n \Delta t$, find the collective potential $\phi_C$ via 
    the methods of Sec. \ref{sec:CollectiveForce}
    \item Calculate density and moment fluxes $f(U^n)$ and source terms $S(U^n)$. 
    \item Obtain $u^{n+1}$ ($t^{n+1} = t + \Delta t $), through the Lax-Friedrichs method
    defined in Sec. \ref{subsec:SourceTermsandSphericalSymmetry}
    \item Set $t\rightarrow t+\Delta t$ and repeat the process   
\end{enumerate*}.

\section{Numerical Results}
\label{sec:results}

For all simulations, we use $N = 7000$ grid points over a domain of length $L = 8$, 
giving a spatial resolution $\Delta r = L / (N - 1) \approx 1.14 \times 10^{-3}$. 
The Courant number $C$, adiabatic index $\gamma$, 
damping parameter $\eta$ (dimensionless), and scattering strength $\Omega_P^2$ are specified for each case.

\subsection{Collective oscillations}

The first step in benchmarking is to verify whether the numerical equilibrium profiles, denoted by $\rho_{0,\text{num}}$,
match the theoretical ones, $\rho_{0,\text{theo}}$.
The latter are computed by numerically solving the equilibrium equation with high resolution.
For each simulation, the system is evolved over a long time interval ($35, \omega_0^{-1}$),
starting from arbitrary initial conditions but ensuring that the (dimensionless) central density is 1\footnote{This can be achieved by normalizing the initial density profile to match the expected (dimensionless) number of particles for the given parameters.}.
As an additional benchmark, we investigate low-frequency breathing modes to compare with theoretical predictions.
For simplicity, we use equilibrium profiles as initial conditions. 
Each profile is unique to parameters \(\gamma\), \(\Omega_P^2\), and \(\eta\), 
determining the particle number. Collective oscillations are 
excited by abruptly changing parameters — in our case, only \(\Omega_P^2\), 
keeping \(\gamma_i = \gamma\) and \(\eta_i = \eta\), with \(\Omega_P^2 < \Omega_{P\,i}^2\) 
to suppress unwanted high-frequency waves. We track the radius of the cloud, $R_{\text{cloud}}(t)$
defined as the sphere containing \(95\%\) of the mass, 
and define the oscillation as \(\Delta R = R_{\text{cloud}}(t) - R\), 
where \(R\) is the final radius. 
Alternatively, we monitor density variations 
\(\Delta \rho\) at a fixed radius \(r_{\text{fixed}} \approx x_0\). 
Each simulation runs for \(T = 35\,\omega_0^{-1}\), outputting data of density, velocity, etc.
every \(N_\text{output}\) iterations.

The oscillation frequencies of $\Delta R$ and $\Delta \rho$ are obtained via
a Discrete Fourier Transform (DFT), using the \texttt{fft} function from
\texttt{numpy}. Prior to the DFT, the data is trimmed to include only the 
region between the second and last peaks. The resulting spectra, $\widehat{\Delta R}(\omega_k)$
and $\widehat{\Delta \rho}(\omega_k)$, are functions of the discrete angular frequencies $\omega_k$.

\begin{figure}[t!]
    \centering
    
    \begin{subfigure}[b]{0.4\textwidth}
        \centering
        \includegraphics[width=\linewidth]{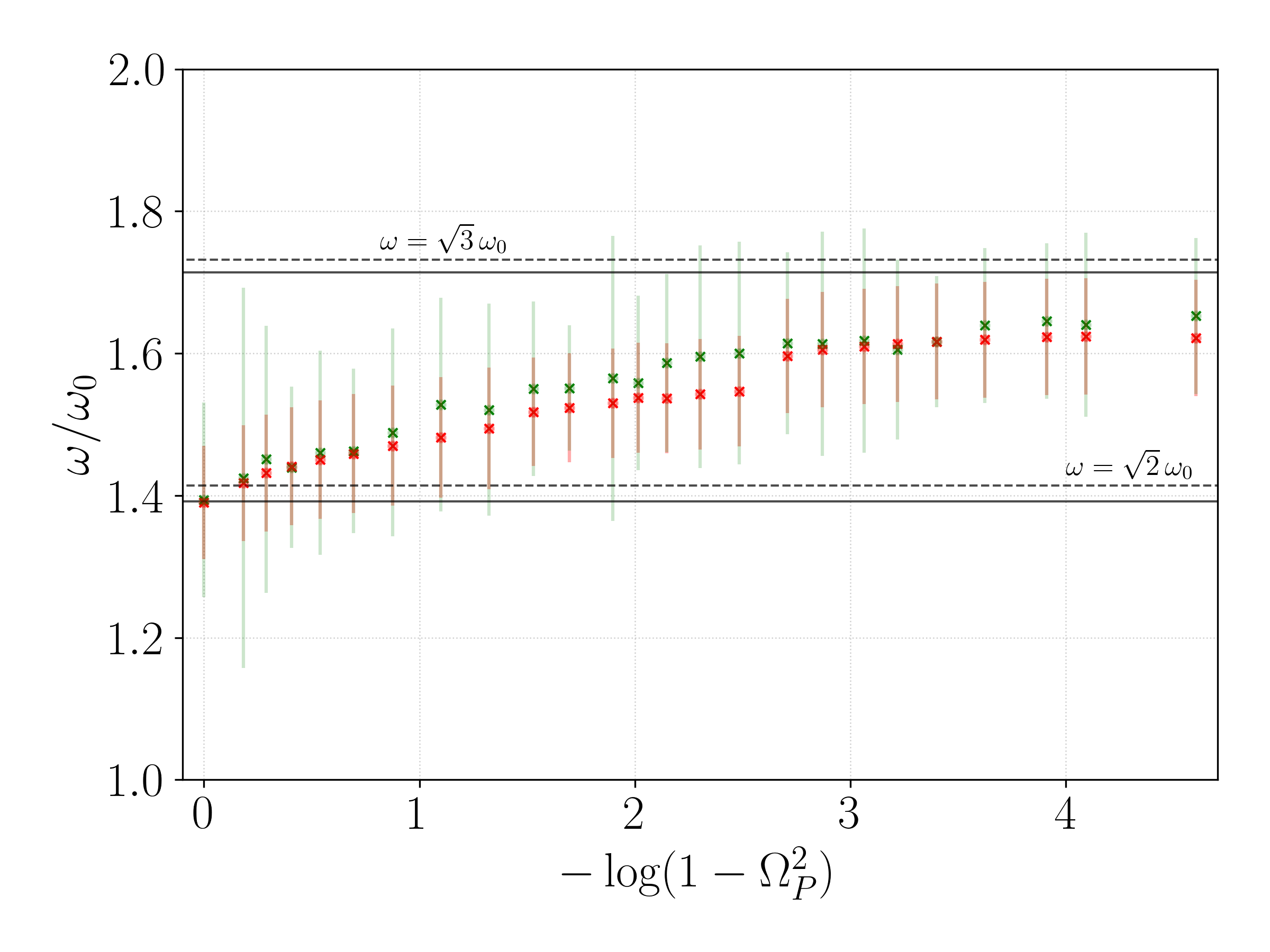}
        \caption{$l=0.5$}
    \end{subfigure}

    \begin{subfigure}[b]{0.4\textwidth}
        \centering
        \includegraphics[width=\linewidth]{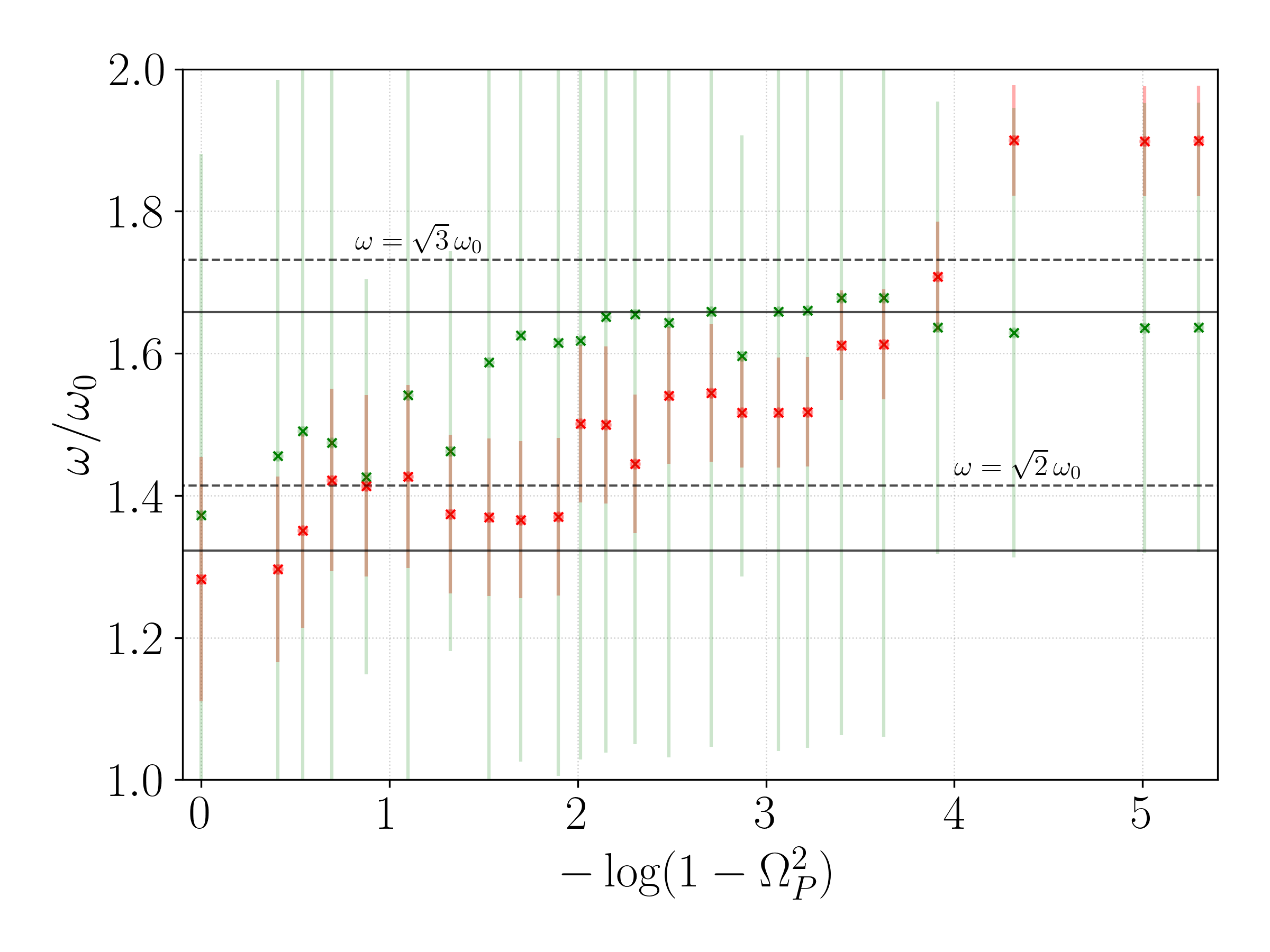}
        \caption{$l=1$}
    \end{subfigure}

    \caption{
        Numerical normal mode frequencies $\omega_{B\,\text{num}}$ from radial oscillations (green) and from density oscillations at $x = 1$ (red), for varying $l$ and fixed $\gamma = 1.5$, $C = 0.25$. 
        Solid lines show known limits: $\omega_B = \sqrt{2}$ (isothermal) and $\omega_B = \sqrt{3}$ (multiple scattering); dashed lines include damping.
    }
    \label{fig:combined_normal_modes}
\end{figure}

To improve accuracy, the breathing mode frequency is computed as a weighted average over nearby frequencies instead of a single peak
($\widehat{\Delta R}(\omega_p)$ or $\widehat{\Delta \rho}(\omega_p)$):
\begin{equation}
    \omega_{B \,\text{num}} = \sum_{j=-N_{\omega}}^{N_{\omega}} \frac{\widehat{F}(\omega_{p - j})}
    {\sum_{i=-N_{\omega}}^{N_{\omega}} \widehat{F}(\omega_{p - i})} \, \omega_{p - j}.
\end{equation}
with the error estimated by the corresponding standard deviation.
We use $N_{\omega} = 2$ throughout. To enhance frequency resolution, we pad 
the time series with $N_\text{output}$ zeros, assuming oscillations vanish beyond the simulation.
The resulting spectra are shown in \Cref{fig:combined_normal_modes}, demonstrating excellent 
agreement with the isothermal limit for all tested $\eta$, though the exact value of the multiple 
scattering limit is not fully reached.

\begin{figure}[t!]
    \centering  
    \includegraphics[width=0.38\textwidth]{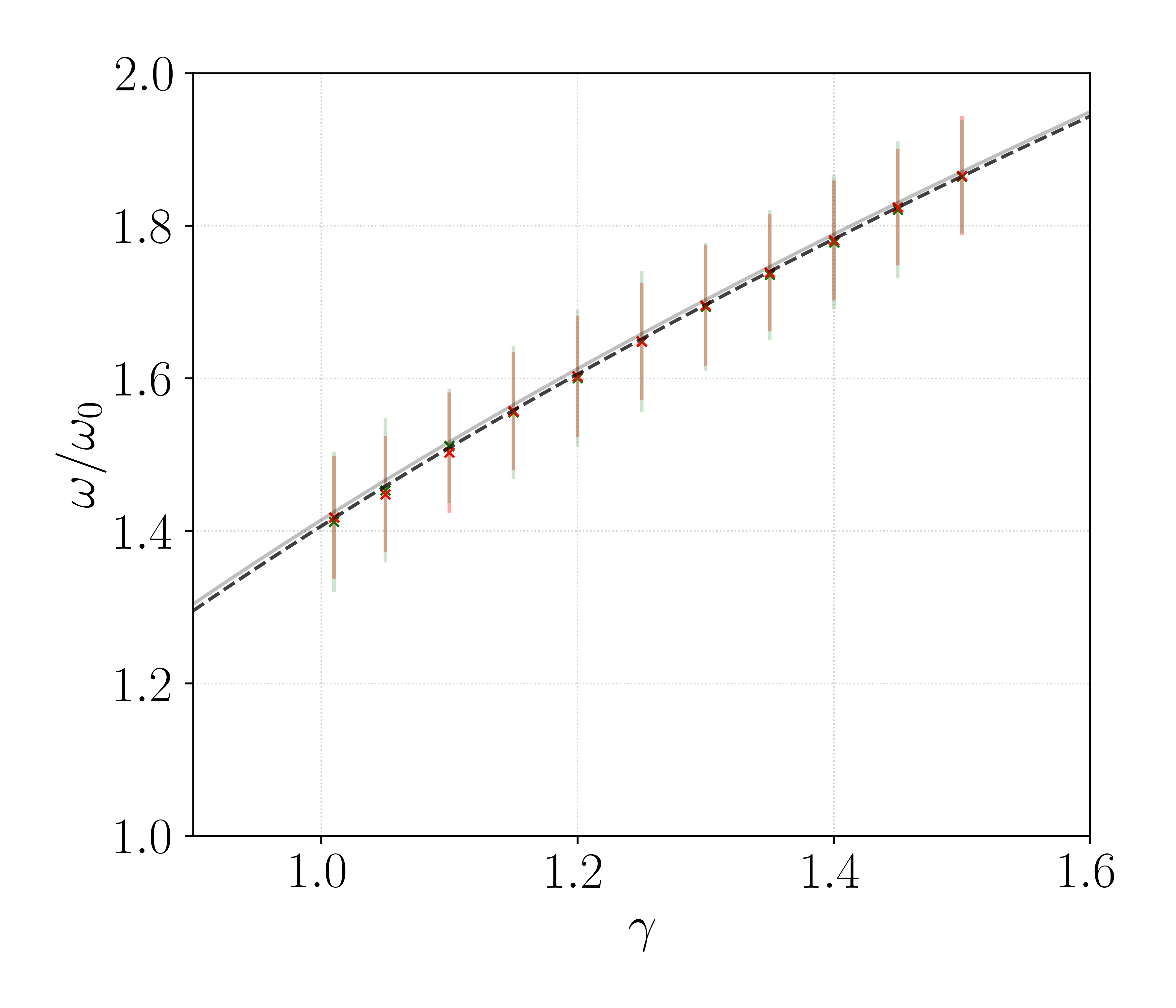}
    \caption{$\omega_{B\,\text{num}}$ from radial oscillations (green) and density oscillations at $x = 1$ 
    (red), for $l=0.3$, $C = 0.25$, $\Omega_P^2 = 0$, $\Omega_{P\,i} = 0.5$, and varying $\gamma$. 
    Dashed and solid lines show analytical results with and without damping.}
    \label{Fig:l0.30.25GammaNormalModes}
\end{figure}

\begin{figure}[t!]
    \centering  
    \includegraphics[width=0.38\textwidth]{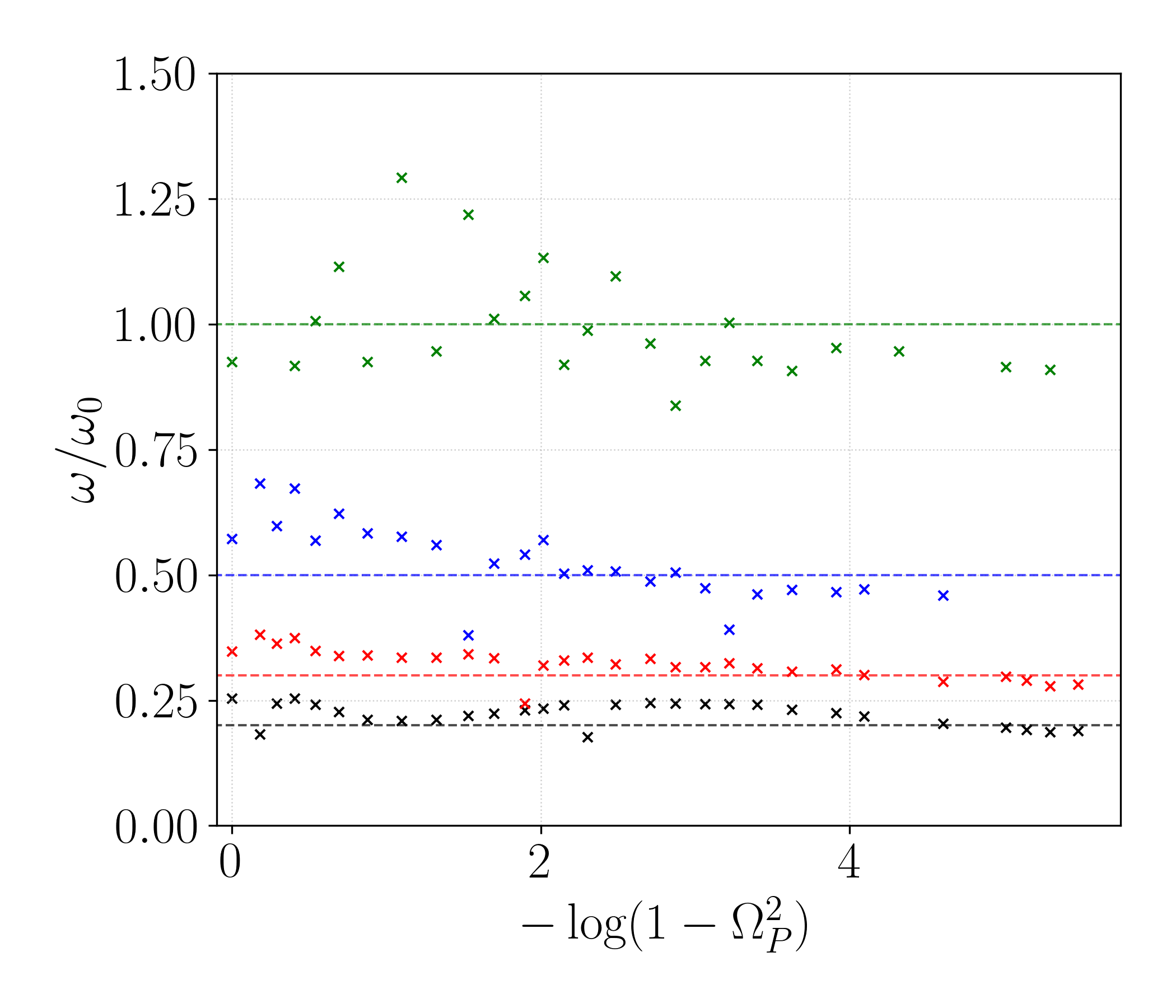}
    \caption{Numerical damping coefficients $\eta_{\text{num}}$ from radial oscillations for various
    $l$ ($l=0.2$ (black)), with $\gamma = 1.5$ and $C = 0.25$.}
    \label{fig:combined_damping_factors_line}
\end{figure}

This behavior likely stems from $\Omega_P^2 = 1$ being a limiting case, which cannot be fully simulated. 
Values beyond $\Omega_P^2 \approx 0.996$ are also challenging, 
as the cloud radius exceeds the domain size. 
Overall, the system transitions from a regime where the 
breathing mode frequency approaches the isothermal 
limit to a saturated high-frequency regime.
As $\eta$ increases, the error in determining the breathing mode frequency also grows.
Higher damping leads to faster equilibration, shortening the excitation window. 
Mitigating this requires a larger drop from $\Omega_{P,i}^2$ to $\Omega_P^2$, though this 
risks exciting higher modes. For varying $\gamma$, \Cref{Fig:l0.30.25GammaNormalModes}
shows good agreement between numerical results and analytical predictions 
(\Cref{NormalModesGamma1Radial}, ~\cite{PhysRevA.88.023412}).
We also estimate the damping coefficient $\eta_{\text{num}}$ by fitting 
the peaks of $\Delta R(t)$ to $A_\eta \exp(-\eta_{\text{num}} t)$ using non-linear least squares. 
Results are shown in \Cref{fig:combined_damping_factors_line}.

\begin{figure}[t]
    \centering

    \begin{subfigure}[b]{0.4\textwidth}
        \centering
        \includegraphics[width=\linewidth]{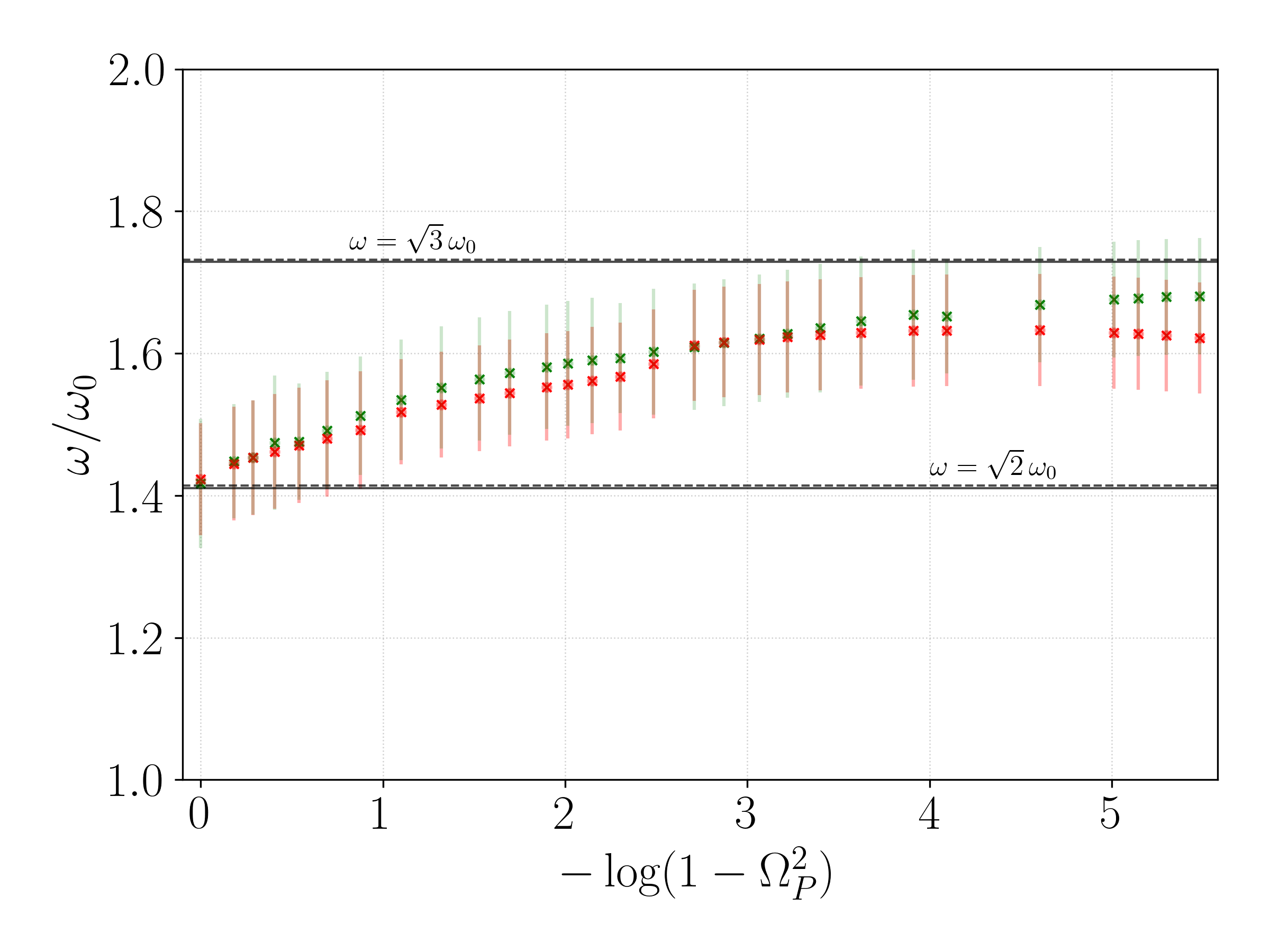}
        \caption{$l=0.2$}
    \end{subfigure}

    \begin{subfigure}[b]{0.4\textwidth}
        \centering
        \includegraphics[width=\linewidth]{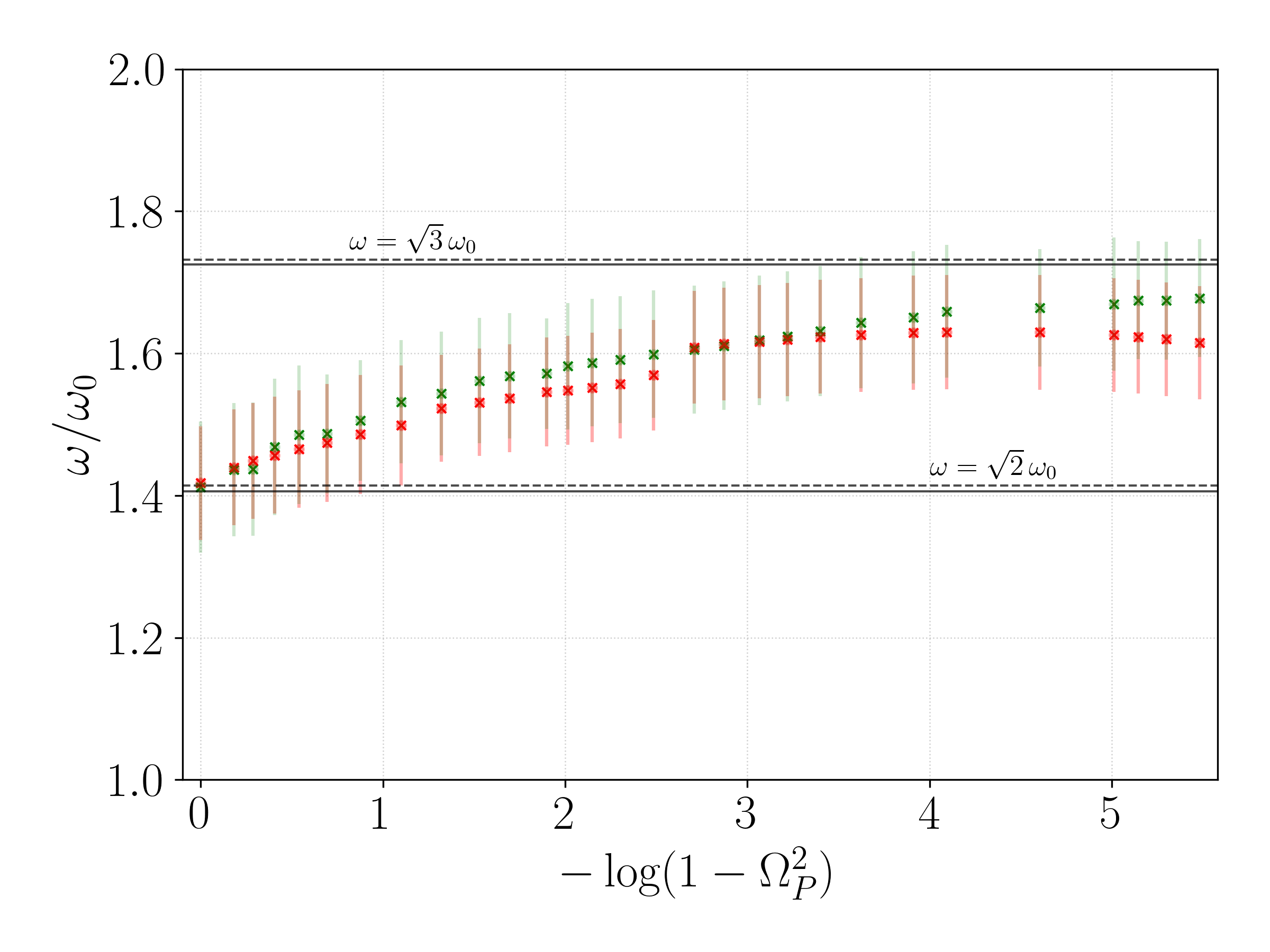}
        \caption{$l=0.3$}
    \end{subfigure}
    \caption{
        Numerical normal mode frequencies $\omega_{B\,\text{num}}$ from radial oscillations (green) and from density oscillations at $x = 1$ (red), for varying $l$ and fixed $\gamma = 1.5$, $C = 0.25$. 
        Solid lines show known limits: $\omega_B = \sqrt{2}$ (isothermal) and $\omega_B = \sqrt{3}$ (multiple scattering); dashed lines include damping.
    }
    \label{fig:combined_normal_modesA}
\end{figure}

\subsection{Shock waves}
\label{sec:shockwaves}

Having now finished some basic precision tests, we move on to
more complex phenomena, for which reference analytical results are more scarce.
In order to simulate shock waves, we shall use a similar strategy to the last section,
though instead of going from $\Omega_{P\,i}^2 > \Omega_P^2$ to $\Omega_P^2$, we shall
proceed the opposite way, i.e the initial density distribution corresponds to 
the equilibrium distribution for $\Omega_{P\,i}^2 <\Omega_P^2$ (considering 
all other parameters constant). If this change is done very quickly enough, the cloud
will rapidally expand and, after reaching a point of maximum expansion (the radius reaches
its peak value), the cloud will then contract, as if rebounding, causing a noticible wave to appear
moving towards the center of the cloud. 

To track the wave, we monitor two key density points: the minimum ($\rho_1$, trough) and maximum 
($\rho_2$, peak). Their positions, $x_1$ and $x_2$, 
evolve in time. We also track the gas velocities at these points and compute the local sound speed 
using $c_s^2 = \gamma \rho^{\gamma-1}$ (in $v_0$ units), denoted $c_{s\,1}$ and $c_{s\,2}$ at 
the peak and trough, respectively.
To estimate the shockwave velocity, we assume it lies midway between 
$x_1$ and $x_2$, and define $x_{\text{shockwave}} = (x_1 + x_2)/2$. 
Over a time window $[t_{\text{min}}, t_{\text{max}}]$, selected 
by inspection, we fit $x_{\text{shockwave}}$ to a quadratic form:
\begin{equation}
    x_{\text{shockwave}}^{\text{fit}} = a_x + b_x (t - t_{\text{min}}) + c_x (t - t_{\text{min}})^2,
\end{equation}
yielding the shockwave velocity as
\begin{equation}
    v_{\text{shockwave}} = b_x + 2 c_x(t - t_{\text{min}}).
\end{equation}
We then compute velocities in the shockwave frame: $v_1 \rightarrow v_1 - v_{\text{shockwave}}$, $v_2 \rightarrow v_2 
- v_{\text{shockwave}}$. As described in Section \ref{sec:shockwaves}, a shock forms if $v_2 > c_{s\,2}$ 
and $v_1 < c_{s\,1}$ in the shock frame. 

\begin{figure*}[t]
    \centering
    \begin{subfigure}{0.32\textwidth}
        \centering
        \includegraphics[width=\linewidth]{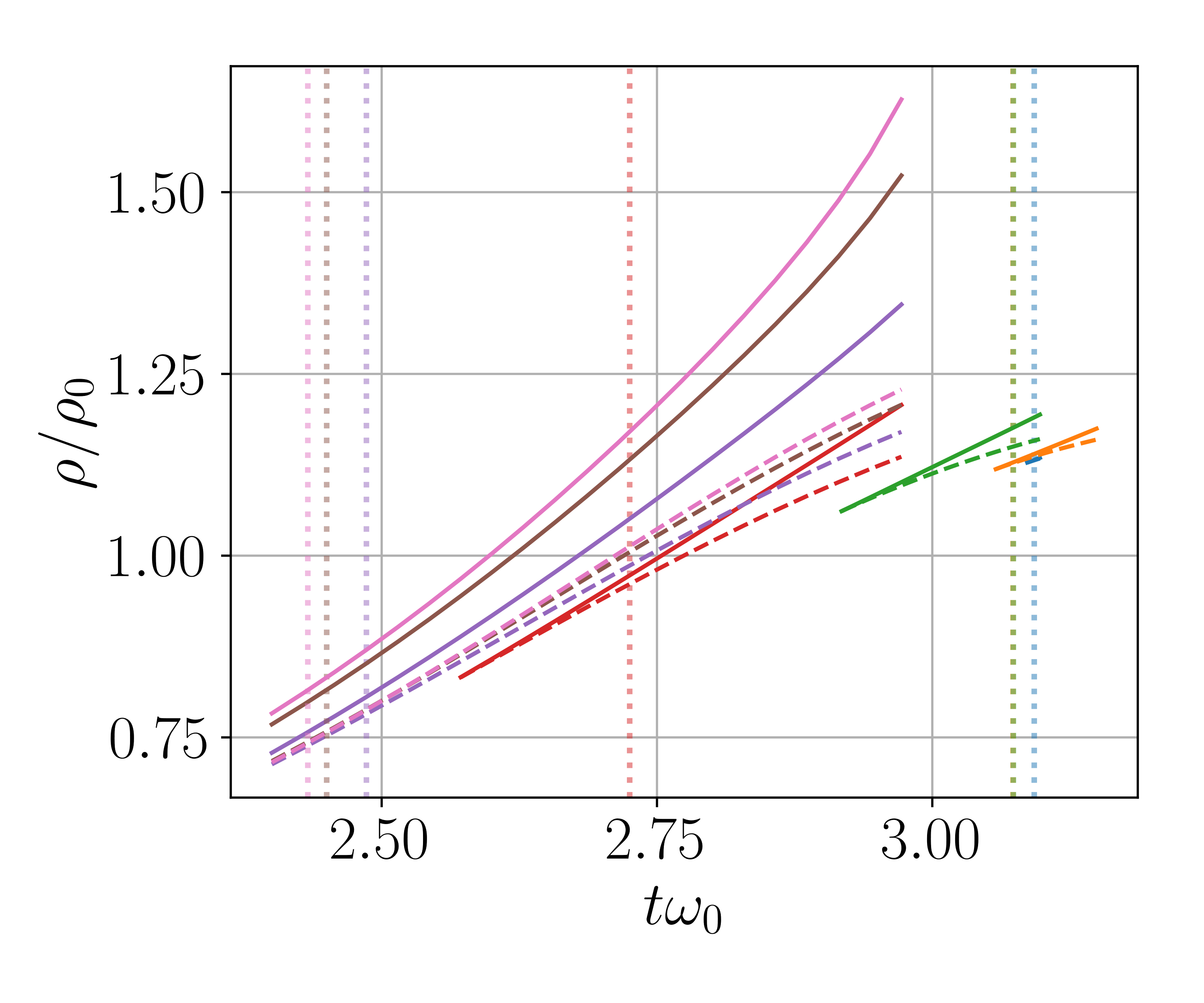}
        \caption{Density evolution of $\rho_1$ (solid) and $\rho_2$ (dashed) for various $\Omega_P^2$.}
        \label{fig:density_evolution}
    \end{subfigure}
    \hfill
    \begin{subfigure}{0.32\textwidth}
        \centering
        \includegraphics[width=\linewidth]{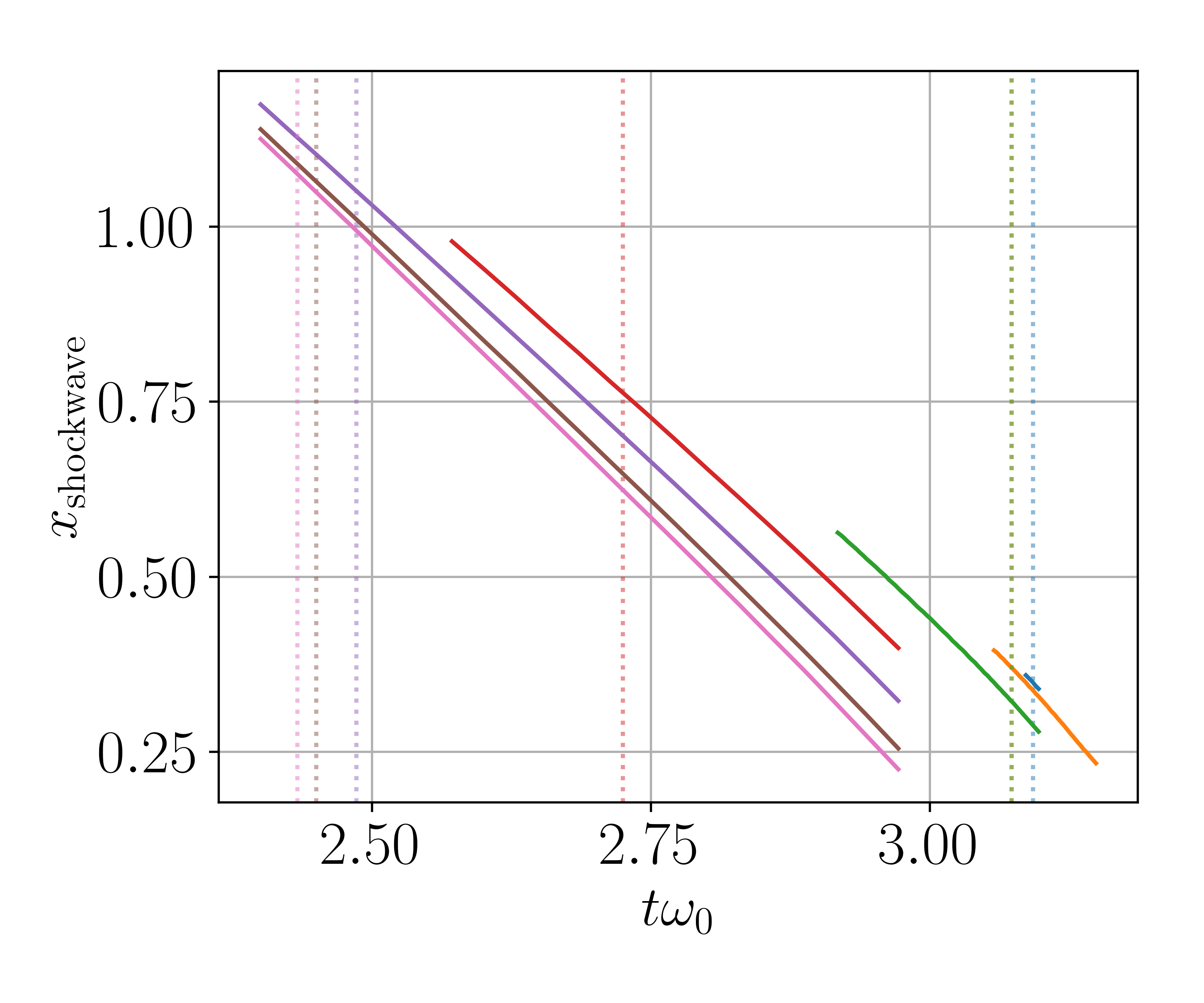}
        \caption{Shockwave position $x_{\mathrm{shockwave}}$ vs. time.}
        \label{fig:position}
    \end{subfigure}
    \hfill
    \begin{subfigure}{0.32\textwidth}
        \centering
        \includegraphics[width=\linewidth]{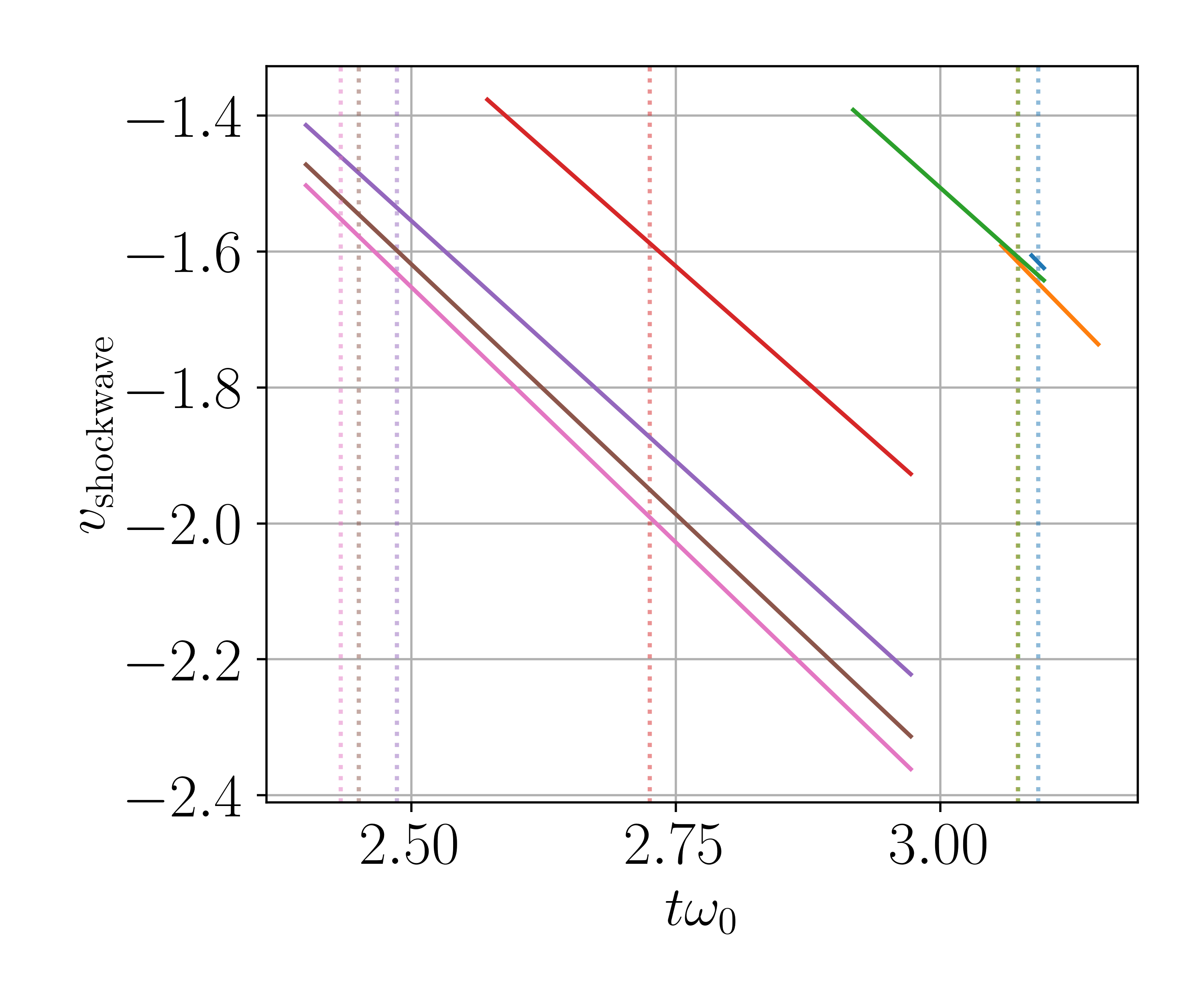}
        \caption{Shockwave velocity $v_{\mathrm{shockwave}}$ over time.}
        \label{fig:velocity}
    \end{subfigure}

    \caption{
        Shockwave formation for $c=0.25$, $\lambda=0.5$, and various $\Omega_P^2$: 
        $0.360$ (blue), $0.366$ (orange), $0.4$ (green), $0.467$ (red), 
        $0.533$ (purple), $0.6$ (brown), $0.633$ (pink). Dashed lines mark shock formation time.
    }
    \label{fig:shockwave_overview}
\end{figure*}

Repeating this 
for various $\Omega_P^2$, we find a threshold $\Omega_{P\,\text{threshold}}^2 \approx 0.360$, 
below which no shockwave forms (see \Cref{fig:shockwave_overview}).

\section{Conclusions}
\label{sec:concl}
In this paper, we developed numerical methods as a foundation for deeper analytical and experimental studies of the Magneto-Optic Trap (MOT). Starting from the basic physics of laser cooling and magnetic trapping, we examined the forces that confine and cool neutral atoms, highlighting their dependence on experimentally tunable parameters. Treating the ultracold atoms as a fluid system governed by continuity and Euler equations incorporating MOT-specific forces, we analyzed equilibrium profiles and normal modes across different regimes.
Building on this framework, we implemented flexible numerical methods that successfully reproduce theoretical expectations in benchmark tests, demonstrating their reliability. As an initial application, we simulated the formation of shock waves triggered by rapid expansion and subsequent contraction of the atomic gas.
For future work, it will be important to extend these methods to more complex phenomena such as shock wave formation and analogues of gravitational collapse. This will require generalizing to other parameter regimes and incorporating additional physics, such as energy dynamics and angular motion. Our current work does not include a convergence analysis, and the artificial diffusion inherent in the methods may limit resolution of high-frequency phenomena. Thus, enhancing numerical accuracy will be essential for studying finer instabilities and modes.
This study lays the groundwork for further exploration of MOT dynamics and their potential use as physical simulators for complex systems.

\bibliography{paperbib}

\begin{thebibliography}{23}%
\makeatletter
\providecommand \@ifxundefined [1]{%
 \@ifx{#1\undefined}
}%
\providecommand \@ifnum [1]{%
 \ifnum #1\expandafter \@firstoftwo
 \else \expandafter \@secondoftwo
 \fi
}%
\providecommand \@ifx [1]{%
 \ifx #1\expandafter \@firstoftwo
 \else \expandafter \@secondoftwo
 \fi
}%
\providecommand \natexlab [1]{#1}%
\providecommand \enquote  [1]{``#1''}%
\providecommand \bibnamefont  [1]{#1}%
\providecommand \bibfnamefont [1]{#1}%
\providecommand \citenamefont [1]{#1}%
\providecommand \href@noop [0]{\@secondoftwo}%
\providecommand \href [0]{\begingroup \@sanitize@url \@href}%
\providecommand \@href[1]{\@@startlink{#1}\@@href}%
\providecommand \@@href[1]{\endgroup#1\@@endlink}%
\providecommand \@sanitize@url [0]{\catcode `\\12\catcode `\$12\catcode `\&12\catcode `\#12\catcode `\^12\catcode `\_12\catcode `\%12\relax}%
\providecommand \@@startlink[1]{}%
\providecommand \@@endlink[0]{}%
\providecommand \url  [0]{\begingroup\@sanitize@url \@url }%
\providecommand \@url [1]{\endgroup\@href {#1}{\urlprefix }}%
\providecommand \urlprefix  [0]{URL }%
\providecommand \Eprint [0]{\href }%
\providecommand \doibase [0]{https://doi.org/}%
\providecommand \selectlanguage [0]{\@gobble}%
\providecommand \bibinfo  [0]{\@secondoftwo}%
\providecommand \bibfield  [0]{\@secondoftwo}%
\providecommand \translation [1]{[#1]}%
\providecommand \BibitemOpen [0]{}%
\providecommand \bibitemStop [0]{}%
\providecommand \bibitemNoStop [0]{.\EOS\space}%
\providecommand \EOS [0]{\spacefactor3000\relax}%
\providecommand \BibitemShut  [1]{\csname bibitem#1\endcsname}%
\let\auto@bib@innerbib\@empty
\bibitem [{\citenamefont {Phillips}(2002)}]{phillips2002nobel}%
  \BibitemOpen
  \bibfield  {author} {\bibinfo {author} {\bibfnamefont {W.~D.}\ \bibnamefont {Phillips}},\ }\bibfield  {title} {\bibinfo {title} {Nobel lecture},\ }in\ \href@noop {} {\emph {\bibinfo {booktitle} {Nobel Lectures, Physics 1996--2000}}},\ \bibinfo {editor} {edited by\ \bibinfo {editor} {\bibfnamefont {G.}~\bibnamefont {Ekspong}}}\ (\bibinfo  {publisher} {World Scientific Publishing Co.},\ \bibinfo {address} {Singapore},\ \bibinfo {year} {2002})\BibitemShut {NoStop}%
\bibitem [{\citenamefont {Migdall}\ \emph {et~al.}(1985)\citenamefont {Migdall}, \citenamefont {Prodan}, \citenamefont {Phillips}, \citenamefont {Bergeman},\ and\ \citenamefont {Metcalf}}]{PhysRevLett.54.2596}%
  \BibitemOpen
  \bibfield  {author} {\bibinfo {author} {\bibfnamefont {A.~L.}\ \bibnamefont {Migdall}}, \bibinfo {author} {\bibfnamefont {J.~V.}\ \bibnamefont {Prodan}}, \bibinfo {author} {\bibfnamefont {W.~D.}\ \bibnamefont {Phillips}}, \bibinfo {author} {\bibfnamefont {T.~H.}\ \bibnamefont {Bergeman}},\ and\ \bibinfo {author} {\bibfnamefont {H.~J.}\ \bibnamefont {Metcalf}},\ }\href@noop {} {\bibfield  {journal} {\bibinfo  {journal} {Phys. Rev. Lett.}\ }\textbf {\bibinfo {volume} {54}},\ \bibinfo {pages} {2596} (\bibinfo {year} {1985})}\BibitemShut {NoStop}%
\bibitem [{\citenamefont {Chu}\ \emph {et~al.}(1986)\citenamefont {Chu}, \citenamefont {Bjorkholm}, \citenamefont {Ashkin},\ and\ \citenamefont {Cable}}]{PhysRevLett.57.314}%
  \BibitemOpen
  \bibfield  {author} {\bibinfo {author} {\bibfnamefont {S.}~\bibnamefont {Chu}}, \bibinfo {author} {\bibfnamefont {J.~E.}\ \bibnamefont {Bjorkholm}}, \bibinfo {author} {\bibfnamefont {A.}~\bibnamefont {Ashkin}},\ and\ \bibinfo {author} {\bibfnamefont {A.}~\bibnamefont {Cable}},\ }\href@noop {} {\bibfield  {journal} {\bibinfo  {journal} {Phys. Rev. Lett.}\ }\textbf {\bibinfo {volume} {57}},\ \bibinfo {pages} {314} (\bibinfo {year} {1986})}\BibitemShut {NoStop}%
\bibitem [{\citenamefont {Chu}\ \emph {et~al.}(1985)\citenamefont {Chu}, \citenamefont {Hollberg}, \citenamefont {Bjorkholm}, \citenamefont {Cable},\ and\ \citenamefont {Ashkin}}]{PhysRevLett.55.48}%
  \BibitemOpen
  \bibfield  {author} {\bibinfo {author} {\bibfnamefont {S.}~\bibnamefont {Chu}}, \bibinfo {author} {\bibfnamefont {L.}~\bibnamefont {Hollberg}}, \bibinfo {author} {\bibfnamefont {J.~E.}\ \bibnamefont {Bjorkholm}}, \bibinfo {author} {\bibfnamefont {A.}~\bibnamefont {Cable}},\ and\ \bibinfo {author} {\bibfnamefont {A.}~\bibnamefont {Ashkin}},\ }\href@noop {} {\bibfield  {journal} {\bibinfo  {journal} {Phys. Rev. Lett.}\ }\textbf {\bibinfo {volume} {55}},\ \bibinfo {pages} {48} (\bibinfo {year} {1985})}\BibitemShut {NoStop}%
\bibitem [{\citenamefont {Haas}\ and\ \citenamefont {Soares}(2022)}]{atoms10030083}%
  \BibitemOpen
  \bibfield  {author} {\bibinfo {author} {\bibfnamefont {F.}~\bibnamefont {Haas}}\ and\ \bibinfo {author} {\bibfnamefont {L.~G.~F.}\ \bibnamefont {Soares}},\ }\href@noop {} {\bibfield  {journal} {\bibinfo  {journal} {Atoms}\ }\textbf {\bibinfo {volume} {10}} (\bibinfo {year} {2022})}\BibitemShut {NoStop}%
\bibitem [{\citenamefont {Sesko}\ \emph {et~al.}(1991)\citenamefont {Sesko}, \citenamefont {Walker},\ and\ \citenamefont {Wieman}}]{SeskoCollectiveForces}%
  \BibitemOpen
  \bibfield  {author} {\bibinfo {author} {\bibfnamefont {D.}~\bibnamefont {Sesko}}, \bibinfo {author} {\bibfnamefont {T.}~\bibnamefont {Walker}},\ and\ \bibinfo {author} {\bibfnamefont {C.}~\bibnamefont {Wieman}},\ }\href@noop {} {\bibfield  {journal} {\bibinfo  {journal} {J. Opt. Soc. Am. B}\ }\textbf {\bibinfo {volume} {8}},\ \bibinfo {pages} {946} (\bibinfo {year} {1991})}\BibitemShut {NoStop}%
\bibitem [{\citenamefont {Dalibard}(1988)}]{DALIBARD1988203}%
  \BibitemOpen
  \bibfield  {author} {\bibinfo {author} {\bibfnamefont {J.}~\bibnamefont {Dalibard}},\ }\href@noop {} {\bibfield  {journal} {\bibinfo  {journal} {Opt. Commun.}\ }\textbf {\bibinfo {volume} {68}},\ \bibinfo {pages} {203} (\bibinfo {year} {1988})}\BibitemShut {NoStop}%
\bibitem [{\citenamefont {Romain}\ \emph {et~al.}(2010)\citenamefont {Romain}, \citenamefont {Hennequin},\ and\ \citenamefont {Verkerk}}]{Romain_2010}%
  \BibitemOpen
  \bibfield  {author} {\bibinfo {author} {\bibfnamefont {R.}~\bibnamefont {Romain}}, \bibinfo {author} {\bibfnamefont {D.}~\bibnamefont {Hennequin}},\ and\ \bibinfo {author} {\bibfnamefont {P.}~\bibnamefont {Verkerk}},\ }\href@noop {} {\bibfield  {journal} {\bibinfo  {journal} {The European Physical Journal D}\ }\textbf {\bibinfo {volume} {61}},\ \bibinfo {pages} {171–180} (\bibinfo {year} {2010})}\BibitemShut {NoStop}%
\bibitem [{\citenamefont {Mendon\ifmmode~\mbox{\c{c}}\else \c{c}\fi{}a}\ \emph {et~al.}(2008)\citenamefont {Mendon\ifmmode~\mbox{\c{c}}\else \c{c}\fi{}a}, \citenamefont {Kaiser}, \citenamefont {Ter\ifmmode~\mbox{\c{c}}\else \c{c}\fi{}as},\ and\ \citenamefont {Loureiro}}]{PhysRevA.78.013408}%
  \BibitemOpen
  \bibfield  {author} {\bibinfo {author} {\bibfnamefont {J.~T.}\ \bibnamefont {Mendon\ifmmode~\mbox{\c{c}}\else \c{c}\fi{}a}}, \bibinfo {author} {\bibfnamefont {R.}~\bibnamefont {Kaiser}}, \bibinfo {author} {\bibfnamefont {H.}~\bibnamefont {Ter\ifmmode~\mbox{\c{c}}\else \c{c}\fi{}as}},\ and\ \bibinfo {author} {\bibfnamefont {J.}~\bibnamefont {Loureiro}},\ }\bibfield  {title} {\bibinfo {title} {Collective oscillations in ultracold atomic gas},\ }\href {https://doi.org/10.1103/PhysRevA.78.013408} {\bibfield  {journal} {\bibinfo  {journal} {Phys. Rev. A}\ }\textbf {\bibinfo {volume} {78}},\ \bibinfo {pages} {013408} (\bibinfo {year} {2008})}\BibitemShut {NoStop}%
\bibitem [{\citenamefont {Terças}(2013)}]{Tercas-Physics-of-Ultra-Cold-Matter}%
  \BibitemOpen
  \bibfield  {author} {\bibinfo {author} {\bibfnamefont {H.~F.~S.}\ \bibnamefont {Terças}},\ }\href@noop {} {}\ (\bibinfo  {publisher} {Springer},\ \bibinfo {year} {2013})\BibitemShut {NoStop}%
\bibitem [{Note1()}]{Note1}%
  \BibitemOpen
  \bibinfo {note} {Throughout this text, partial derivatives are abbreviated as $\partial _{x}$ for clarity and brevity.}\BibitemShut {Stop}%
\bibitem [{\citenamefont {{Kippenhahn}}\ and\ \citenamefont {{Weigert}}(1990)}]{stellarStucEvol}%
  \BibitemOpen
  \bibfield  {author} {\bibinfo {author} {\bibfnamefont {R.}~\bibnamefont {{Kippenhahn}}}\ and\ \bibinfo {author} {\bibfnamefont {A.}~\bibnamefont {{Weigert}}},\ }\href@noop {} {\emph {\bibinfo {title} {"Stellar Structure and Evolution"}}}\ (\bibinfo  {publisher} {Springer},\ \bibinfo {year} {1990})\BibitemShut {NoStop}%
\bibitem [{\citenamefont {Rodrigues}\ \emph {et~al.}(2016)\citenamefont {Rodrigues}, \citenamefont {Rodrigues}, \citenamefont {Moreira}, \citenamefont {Terças},\ and\ \citenamefont {Mendonça}}]{PhysRevA.93.023404}%
  \BibitemOpen
  \bibfield  {author} {\bibinfo {author} {\bibfnamefont {J.~D.}\ \bibnamefont {Rodrigues}}, \bibinfo {author} {\bibfnamefont {J.~A.}\ \bibnamefont {Rodrigues}}, \bibinfo {author} {\bibfnamefont {O.~L.}\ \bibnamefont {Moreira}}, \bibinfo {author} {\bibfnamefont {H.}~\bibnamefont {Terças}},\ and\ \bibinfo {author} {\bibfnamefont {J.~T.}\ \bibnamefont {Mendonça}},\ }\href@noop {} {\bibfield  {journal} {\bibinfo  {journal} {Phys. Rev. A}\ }\textbf {\bibinfo {volume} {93}},\ \bibinfo {pages} {023404} (\bibinfo {year} {2016})}\BibitemShut {NoStop}%
\bibitem [{\citenamefont {Ter\ifmmode~\mbox{\c{c}}\else \c{c}\fi{}as}\ and\ \citenamefont {Mendon\ifmmode~\mbox{\c{c}}\else \c{c}\fi{}a}(2013)}]{PhysRevA.88.023412}%
  \BibitemOpen
  \bibfield  {author} {\bibinfo {author} {\bibfnamefont {H.}~\bibnamefont {Ter\ifmmode~\mbox{\c{c}}\else \c{c}\fi{}as}}\ and\ \bibinfo {author} {\bibfnamefont {J.~T.}\ \bibnamefont {Mendon\ifmmode~\mbox{\c{c}}\else \c{c}\fi{}a}},\ }\href@noop {} {\bibfield  {journal} {\bibinfo  {journal} {Phys. Rev. A}\ }\textbf {\bibinfo {volume} {88}},\ \bibinfo {pages} {023412} (\bibinfo {year} {2013})}\BibitemShut {NoStop}%
\bibitem [{\citenamefont {Dalfovo}\ \emph {et~al.}(1999)\citenamefont {Dalfovo}, \citenamefont {Giorgini}, \citenamefont {Pitaevskii},\ and\ \citenamefont {Stringari}}]{RevModPhys.71.463}%
  \BibitemOpen
  \bibfield  {author} {\bibinfo {author} {\bibfnamefont {F.}~\bibnamefont {Dalfovo}}, \bibinfo {author} {\bibfnamefont {S.}~\bibnamefont {Giorgini}}, \bibinfo {author} {\bibfnamefont {L.~P.}\ \bibnamefont {Pitaevskii}},\ and\ \bibinfo {author} {\bibfnamefont {S.}~\bibnamefont {Stringari}},\ }\href@noop {} {\bibfield  {journal} {\bibinfo  {journal} {Rev. Mod. Phys.}\ }\textbf {\bibinfo {volume} {71}},\ \bibinfo {pages} {463} (\bibinfo {year} {1999})}\BibitemShut {NoStop}%
\bibitem [{\citenamefont {Pinsonneault}\ and\ \citenamefont {Ryden}(2023)}]{Pinsonneault_Ryden_2023}%
  \BibitemOpen
  \bibfield  {author} {\bibinfo {author} {\bibfnamefont {M.}~\bibnamefont {Pinsonneault}}\ and\ \bibinfo {author} {\bibfnamefont {B.}~\bibnamefont {Ryden}},\ }\href@noop {} {}\ (\bibinfo  {publisher} {Cambridge University Press},\ \bibinfo {year} {2023})\BibitemShut {NoStop}%
\bibitem [{\citenamefont {Landau}\ and\ \citenamefont {Lifshitz}(2013)}]{landauVol6}%
  \BibitemOpen
  \bibfield  {author} {\bibinfo {author} {\bibfnamefont {L.}~\bibnamefont {Landau}}\ and\ \bibinfo {author} {\bibfnamefont {E.}~\bibnamefont {Lifshitz}},\ }\href@noop {} {\emph {\bibinfo {title} {"Course of Theoretical Physics"}}},\ \bibinfo {number} {vol. 6}\ (\bibinfo  {publisher} {Elsevier Science},\ \bibinfo {year} {2013})\BibitemShut {NoStop}%
\bibitem [{Note2()}]{Note2}%
  \BibitemOpen
  \bibinfo {note} {This generalizes to multiple dimensions ($\vb {x} \in \protect \mathbb {R}^n$).}\BibitemShut {Stop}%
\bibitem [{\citenamefont {LeVeque}(1992)}]{leveque1992numerical}%
  \BibitemOpen
  \bibfield  {author} {\bibinfo {author} {\bibfnamefont {R.}~\bibnamefont {LeVeque}},\ }\href@noop {} {\emph {\bibinfo {title} {Numerical Methods for Conservation Law}}}\ (\bibinfo  {publisher} {Springer Basel AG},\ \bibinfo {year} {1992})\BibitemShut {NoStop}%
\bibitem [{\citenamefont {Chen}(2001)}]{Chen2001}%
  \BibitemOpen
  \bibfield  {author} {\bibinfo {author} {\bibfnamefont {G.-Q.}\ \bibnamefont {Chen}},\ }\href@noop {} {\bibfield  {journal} {\bibinfo  {journal} {Acta Math. Univ. Comenianae}\ }\textbf {\bibinfo {volume} {70}},\ \bibinfo {pages} {51} (\bibinfo {year} {2001})}\BibitemShut {NoStop}%
\bibitem [{Note3()}]{Note3}%
  \BibitemOpen
  \bibinfo {note} {The asterisks have been dropped, for the sake of simplity}\BibitemShut {NoStop}%
\bibitem [{Note4()}]{Note4}%
  \BibitemOpen
  \bibinfo {note} {The convention used in this paper is $\protect \hat {f}(\vb {k}) = \DOTSI \intop \ilimits@ e^{-2 \pi i \vb {k} \cdot \vb {x}} f (\vb {x})\protect \,\protect \text {d}\vb {x}^{3}$ for the FT and $f(\vb {k}) = \DOTSI \intop \ilimits@ e^{2 \pi i \vb {k} \cdot \vb {x}} \protect \hat {f} (\vb {k})\protect \,\protect \text {d}\vb {k}^{3}$ for the IFT}\BibitemShut {NoStop}%
\bibitem [{Note5()}]{Note5}%
  \BibitemOpen
  \bibinfo {note} {This can be achieved by normalizing the initial density profile to match the expected (dimensionless) number of particles for the given parameters.}\BibitemShut {Stop}%
\end{thebibliography}%

\end{document}